\documentstyle[12pt,epsfig]{article}

\newcommand{\AmS}{{\protect\the\textfont2
  A\kern-.1667em\lower.5ex\hbox{M}\kern-.125emS}}

\newcommand{\be}{\begin{equation}}
\newcommand{\ee}{\end{equation}}
\def\15{\raise.9ex\hbox{1-5)\kern-0.47em}\kern.4em}
\def\69{\raise.9ex\hbox{6-9)\kern-0.47em}\kern.4em}

\begin{document}
\newcommand{\reaction}{ $\nu_{\tau}$N$\rightarrow$$\tau^{-}$ X} 
\newcommand{\decay}{$\tau^{-}\rightarrowe^{-}+\nu_{\tau}+\bar\nu_{e}$}
\newcommand{\osc}{ $\nu_{\mu}\rightarrow\nu_{\tau}$}        
\newcommand{\dm}{$\Delta m^{2}$}
\begin{center}
\centerline{ \bf MUON ENERGY ESTIMATE THROUGH MULTIPLE}
\centerline{\bf SCATTERING WITH THE MACRO DETECTOR}
\vskip 1cm
\end{center}
\nobreak\bigskip\nobreak
\pretolerance=10000
\bigskip
\begin{center}
{\bf The MACRO Collaboration}\\
\nobreak\bigskip\nobreak
M.~Ambrosio$^{12}$, 
R.~Antolini$^{7}$, 
G.~Auriemma$^{14,a}$, 
D.~Bakari$^{2,17}$, 
A.~Baldini$^{13}$, 
G.~C.~Barbarino$^{12}$, 
B.~C.~Barish$^{4}$, 
G.~Battistoni$^{6,b}$, 
Y.~Becherini$^{2}$,
R.~Bellotti$^{1}$
% ...
\\
\vspace{1.5 cm}
\newpage
\footnotesize
1. Dipartimento di Fisica dell'Universit\`a  di Bari and INFN, 70126 Bari, 
Italy \\
2. Dipartimento di Fisica dell'Universit\`a  di Bologna and INFN, 40126 
Bologna, Italy \\
3. Physics Department, Boston University, Boston, MA 02215, USA \\
4. California Institute of Technology, Pasadena, CA 91125, USA \\
5. Department of Physics, Drexel University, Philadelphia, PA 19104, USA \\
6. Laboratori Nazionali di Frascati dell'INFN, 00044 Frascati (Roma), Italy \\
7. Laboratori Nazionali del Gran Sasso dell'INFN, 67010 Assergi (L'Aquila), 
Italy \\
8. Depts. of Physics and of Astronomy, Indiana University, Bloomington, 
IN 47405, USA \\
9. Dipartimento di Fisica dell'Universit\`a  dell'Aquila and INFN, 67100 
L'Aquila, Italy\\
10. Dipartimento di Fisica dell'Universit\`a  di Lecce and INFN, 73100 Lecce, 
Italy \\
11. Department of Physics, University of Michigan, Ann Arbor, MI 48109, 
USA \\
12. Dipartimento di Fisica dell'Universit\`a  di Napoli and INFN, 80125 
Napoli, Italy \\
13. Dipartimento di Fisica dell'Universit\`a  di Pisa and INFN, 56010 Pisa, 
Italy \\
14. Dipartimento di Fisica dell'Universit\`a  di Roma "La Sapienza" and 
INFN, 00185 Roma, Italy \\
15. Physics Department, Texas A\&M University, College Station, TX 77843, 
USA \\
16. Dipartimento di Fisica Sperimentale dell'Universit\`a  di Torino and 
INFN, 10125 Torino, Italy \\
17. L.P.T.P, Faculty of Sciences, University Mohamed I, B.P. 524 Oujda, 
Morocco \\
18. Dipartimento di Fisica dell'Universit\`a  di Roma Tre and INFN Sezione 
Roma Tre, 00146 Roma, Italy \\
$a$ Also Universit\`a  della Basilicata, 85100 Potenza, Italy \\
$b$ Also INFN Milano, 20133 Milano, Italy \\
$c$ Also Istituto TESRE/CNR, 40129 Bologna, Italy \\
$d$ Also Universit\`a  di Trieste and INFN, 34100 Trieste, Italy \\
$e$ Also Dipartimento di Energetica, Universit\`a  di Roma, 00185 Roma, Italy \\
$f$ Also Institute for Nuclear Research, Russian Academy of Science, 117312 
Moscow, Russia \\
$g$ Also Institute for Space Sciences, 76900 Bucharest, Romania \\
$h$ Macalester College, Dept. of Physics and Astr., St. Paul, MN 55105 \\
$i$ Also Department of Physics, Pennsylvania State University, University 
Park, PA 16801, USA \\
$l $Also Dipartimento di Fisica dell'Universit\`a  della Calabria, Rende 
(Cosenza), Italy \\
$m$ Also Department of Physics, James Madison University, Harrisonburg, 
VA 22807, USA \\
$n$ Also RPD, PINSTECH, P.O. Nilore, Islamabad, Pakistan \\
$o$ Also Dipartimento di Scienze e Tecnologie Avanzate, Universit\`a  del 
Piemonte Orientale, Alessandria, Italy \\
$p$ Also U. Minn. Duluth Physics Dept., Duluth, MN 55812 \\
$q$ Also Dept. of Physics, MIT, Cambridge, MA 02139 \\
$r$ Also Intervideo Inc., Torrance CA 90505 USA \\
$s$ Also Resonance Photonics, Markham, Ontario, Canada\\
$t$ Now at INFN Bologna, Via Irnerio 46, 40126 Bologna, Italy\\
\newpage
\baselineskip=14.5pt
\begin{abstract}
Muon energy measurement represents an important issue for
any experiment addressing neutrino induced upgoing muon studies.
Since the neutrino oscillation probability depends on the neutrino
energy, a measurement of the muon energy adds an important piece
of information concerning the neutrino system.
We show in this paper how the MACRO limited streamer tube system 
can be operated in drift mode by
using the TDC's included in the QTPs, an electronics
designed for magnetic monopole search. 
An improvement
of the space resolution is obtained, through an analysis of
the multiple scattering of 
muon tracks as they pass through our detector. This 
information can be used further to obtain an estimate
of the energy of muons crossing the detector.
Here we present the results of two dedicated tests, performed at
CERN PS-T9 and SPS-X7 beam lines, to provide a full check of the
electronics and to exploit the feasibility of such a multiple scattering
analysis. We show that by using a neural network approach, we
are able to reconstruct the muon energy for $E_\mu<$40 GeV. The test beam
data provide an absolute energy calibration, which allows us to apply
this method to MACRO data.
\end{abstract}
\end{center}

\baselineskip=17pt

\newpage

\section{Introduction}

The most recent studies of neutrino induced up-going muons have been
performed by two experiments:
Super--Kamiokande\cite{sk}, using a water Cherenkov detector, and MACRO\cite{macroneutrino}, 
tagging neutrino events with a time of flight technique. Both experiments
observed
a flux deficit and a distortion of the up-going muon angular distribution
with respect to the Monte Carlo expectation.
The oscillation probability of neutrinos depends on the oscillation
parameters ($\Delta m^{2}$,$~$$sin^{2}2$$\theta$) and on the ratio L/E, where
L is the distance between neutrino 
production and interaction point, while E is the neutrino energy.
The energy of up-going neutrinos, interacting in the rock below
the apparatus, is shared by the up-going muon and by the hadrons.
Independent of the detector resolution, a precise measurement of the muon energy is 
prevented by the 
energy lost by the muon in the rock, while the hadrons are absorbed in the
rock.
Nevertheless the residual
muon energy can in principle be measured. 
In this paper we explore the possibility
of  performing such a measurement relying on 
muon multiple scattering(MS). 
The r.m.s. of the lateral displacement of the muon trajectory on a 
projected plane of material with depth X
and radiation length $X_{o}$, can 
be written as:
\begin{equation}
\sigma_{proj}^{MS}\simeq 
\frac{X}{\sqrt{3}}
\frac{0.0136} {p\beta c}
\sqrt{\frac{X}{X_{o}}}(1+0.038ln(X/X_{o}))
\end{equation}
where p is in GeV/c and
for MACRO, X$\simeq$25$X_{o}$/cos$\theta$, giving for
vertical muons $\sigma_{proj}^{MS}$$\simeq$10$cm$/E(GeV).

For a given amount of crossed material, the
capability of measuring track deflection is possible only when the particle
displacement due to the multiple scattering is larger than  the detector
space resolution. The space point resolution of the tracking system of
MACRO's (3x3)$cm^{2}$ cross section streamer tubes is of the order of 1 $cm$, 
and therefore
provides a muon energy estimate through MS 
up to $\simeq$ 10GeV. 
Supposing $\Delta m^{2}$=$\cal O$($10^{-3}$$eV^{2}$) and
$sin^{2}2$$\theta$$\simeq$1, 
the neutrino induced up-going muons, are 
not expected to experience neutrino oscillation at all energies. At the 
up-going muon median energy in MACRO, 11 GeV\cite{iomaxbat}, the oscillation probability 
is still as high as 50$\%$(Fig. 1), while it's just 10$\%$ for
$E_{\mu}$=40 GeV:
an improvement of the space resolution 
offers
the possibility of evaluating muon energy over a
sufficiently wide energy range.

In order to achieve
this goal, we retrieve drift time information from the limited streamer tubes by using the TDC's implemented in the MACRO 
QTP electronic system\cite{napoletani}. 

In this paper we describe the application of this electronics
to evaluate the MS effect along a muon track, showing the results obtained
with  
two dedicated tests, performed at CERN PS-T9 and SPS-X7 beam lines, in October 2000
and August 2001 respectively. The application of the method to MACRO
data is then presented.

\section{The MACRO limited streamer tubes in drift mode}
The MACRO streamer tube system\cite{macronim} consists of about 5,600 
chambers; each 
chamber is made of 8 streamer tubes with cross section (3x3)$cm^{2}$ and
1200 $cm$ length, for a total of about 50,000 wires. 
These tubes were
built in ``coverless'' mode, i.e. the electric field of the inner 
four walls is not exactly the same. 
Despite this feature as well as the large cell dimension,
the intrinsic space resolution of these chambers can be quite good,
as demonstrated in (\cite{ioebattistoni}) where using a MACRO streamer 
tube in drift mode, a resolution
of $\sigma$$\simeq$250$\mu m$ was obtained using standard Lecroy 2228A
TDC (0.25 ns/bin). Such resolution has to be considered as the ultimate
resolution achievable with this device.

Although the MACRO streamer tube electronics does not contain a 
high resolution TDC system,
information on streamer timing can be extracted 
using the QTP system\cite{napoletani}.
This electronics, designed for our magnetic monopole search\cite{monopoli},
consists of a ADC/TDC system and acts as a 640 $\mu s$
memory, during which the 
charge, the arrival time and the width of the streamer pulse of the 
particle crossing the cell are recorded.
A slow particle
in MACRO ($\beta$$\geq$$10^{-4}$) may take more than 500~$\mu$s to 
cross the 
detector. The QTP-TDC system allows us
to distinguish randomly distributed background hits in this time window 
from a genuine slow particle, which, during the crossing time
of the detector, describes a line in the space-time plane. 
For the magnetic monopole reconstruction optimization, 
a distributed clock of 20/3 MHz was chosen, resulting in an equivalent
TDC bin width of $\Delta T$$\simeq$150~ns.
This clock frequency is quite coarse
for drift time measurements in a single cell, 
given that the maximum drift time
for MACRO streamer tubes, operated with a He(73$\%$)/n-pentane(27$\%$)
mixture, is $\simeq$600ns.
The ultimate resolution that can be therefore obtained with 
such a system is
$\sigma$$\simeq$$v_{drift}$$\times$$\Delta$T/$\sqrt{12}$$\simeq$1.9mm, 
which is about an order of magnitude greater than the intrinsic precision
of the streamer
tube, operated in drift mode. 
Nevertheless, if such improved resolution could be achieved,
it would 
be sufficient to estimate up-going muon energies up to 30-40 GeV.

In order to reduce the number of electronic channels,
a single MACRO QTP channel, serves the OR of 4 chambers, for a total of
32 wires. 
Selecting only planes  
with a single fired tube, the association
with the fired QTP channel is uniquely determined.

Given that our electronics was not designed for drift time measurements,
the relative linearity was tested
only for the much larger time scale of 500$\mu$s rather 
than 600~ns. To avoid any systematic effects and 
to fully understand the capability of the QTP system in this context, 
we decided to test the electronics in a beam test at CERN PS-T9.
\section{Streamer tube system performance in drift mode}
To study the QTP-TDC's linearity,
the drift velocity in He/n-pentane mixture and to develop
the software used for muon tracking, we performed a test beam run in
CERN PS-T9 beamline in October 2000.

For these tests, we reproduced a slice of the MACRO 
detector using 14 coverless streamer tube 
chambers, (25$\times$3$\times$200)$cm^3$, filled with the standard MACRO
gas mixture. The rock absorbers reproduced
as much as possible those of MACRO. We built 7 iron boxes, 
(40$\times$40$\times$32)$cm^3$, filled
with rock excavated from the Gran Sasso tunnel ($\rho$=2.0 g/$cm^{3}$).
As in MACRO,
each streamer tube chamber was equipped with a streamer tube
read-out card and the analog output of a chamber was sent to 
a QTP channel. The 
digital output, OR of each chamber signals, was sent to a Lecroy 2228A TDC. 
Such double 
measurement of the drift time allowed us to make a comparison 
between QTP-TDC's and Lecroy
TDC's on an event by event basis.
The test beam layout is shown in Fig. 2.  The trigger was provided by
a fast coincidence of the scintillators S1,$~$S2,$~$S3. The last scintillator,
following a 60 $cm$ iron slab, 
suppresses the $\pi$,$K$ contamination in the beam at high energies.
The data acquisition was performed using LabView, running 
on a MacIntosh Quadra 950.
Fig. 3 shows the plateau curve of the streamer tubes used in the test beam.
We operated these chambers at HV=4050~V, where a full efficiency is reached.
We collected 60 runs, with the beam stoppers closed, for a total of
about $10^{5}$ muons, 
with energy ranging from
2 to 12 GeV. Several runs were also taken with the rock absorbers removed, 
to study the QTP electronics and to allow for space resolution
evaluation, without contributions of multiple scattering in the absorbers
at these low muon energies.

First, we evaluated the QTP-TDC's linearity, by comparing its
data with that recorded by the Lecroy TDC's.
Fig. 4 shows the relationship between these
two measurements, for values of the 
QTP-TDC system(75~ns, 225~ns, 375~ns, 525~ns), where we took 
the average of the Lecroy TDC's time distribution. The errors
represents the width of the QTP-TDC's and the rms of the
corresponding Lecroy-TDC time distributions. 
 
Although the maximum drift time in our streamer tubes is about 600 ns, 
due to the non-homogeneity of the electric field in the streamer tube 
cell\cite{ioebattistoni},
the region between 500 ns$\leq$T$\leq$600 ns 
is not uniformly populated.
We evaluated that this effect accounts
for the $\simeq$10$\%$ observed shift-up of the QTP-TDCs, with respect to the expected 
average in that bin.

For T$\leq$450 ns there is full consistency with Lecroy-TDC measurement.
Considering the coarseness of QTP-TDC we conclude that the
comparison if fully satisfactory. Therefore
we used the central value of each QTP-TDC bin (150 ns wide).

We then studied the drift velocity in He/n-pentane mixture. Since in the 
test beam configuration the N muons hit the 
detector at normal incidence:
\begin{equation}
\frac{dN}{dt}=\frac{dN}{dx}\frac{dx}{dt}=\frac{dN}{dx}\cdot v_{drift}=
K\cdot v_{drift}.
\end{equation} 
The evaluation of $v_{drift}$ can be therefore obtained fitting the Lecroy 
TDC spectrum
distribution.  

Fig. 5 shows the experimental results obtained,
where we have superimposed
the results of a GARFIELD\cite{garfield} simulation for comparison.
Such code performs a detailed simulation of electron drift and signal 
generation in gaseous wire detectors. We described the electrostatic
structure of a limited streamer tube (an anode wire at the center of a
square cross section cathode) by means of a lattice made of 81 wires,
spaced by 3 cm, kept at the proper voltage (alternating the sign): 
the central cell in this lattice corresponds to the actual cell. 
The drift velocity as a function of electric field has been computed
assuming the standard MACRO gas mixture by using the GARFIELD-MagBoltz
interface.
The experimental data are in agreement with the simulation.
 
Once the TDC linearity has been checked and the $v_{drift}$ has been measured,
the test beam data can be used to measure the space resolution. 
Fig. 6 shows the residuals distribution for streamer tubes in drift mode using
the Lecroy TDC's and the QTP-TDC system. 
Using the LeCroy TDC data, we find
a resolution of 500 $\mu$m, while for the 
QTP-TDC data we obtained a resolution of $\sigma$$\simeq$2 mm.
This resolution limit is very close
to that expected based on QTP-TDC's time resolution 
$\sigma$=$v_{drift}$$\times$150ns/$\sqrt{12}$$\simeq$
4 $cm$/$\mu s$$\times$150ns/$\sqrt{12}$$\simeq$1.9~mm. 
\section{\bf Study of the MACRO space resolution}

To estimate the performance of the streamer tubes operated in drift mode in
MACRO, 
we analysed a down-going muon sample,
whose average energy is $<$$E_{\mu}$$>$$\simeq$ 320 GeV\cite{macrotrd}.

The analysis was performed by using the following steps:\\
1) We considered the muon track reconstructed with the standard MACRO
tracking (i.e. no QTP information is used at this stage);\\
2) We selected those hits containing only a single fired tube;\\
3) For each hit we looked at the corresponding QTP-TDC value in a time window of
2 $\mu$s. Given the background rate in the MACRO 
streamer tubes, $\simeq$40 Hz/$m^2$,
this corresponds to $\simeq$ 480 Hz on 4 chambers (~1~QTP~channel~), giving 
a probability $\simeq$ $10^{-3}$ for a spurious hit to mimic a genuine
QTP~-~TDC count;\\
4) After converting the TDC values to drift radii, by using the drift
velocity measured in the test beam, a global fit of the track is performed.

As a first step, we used this procedure to perform an alignment
of the detector database. The standard MACRO database was  
computed using the streamer tube data
in digital mode, hence to take advantage of the improved 
space resolution achieved 
by this method, we first had to upgrade the precision 
of the detector database. To accomplish this
we used 15$\times 10^{6}$
down-going muon tracks. Since the MACRO streamer 
tubes, 1200 $cm$ long, are made
of PVC, a flexible material, part of the 
misalignment may come from the deviation
from a straight line along
the main axis of each streamer tube (sagitta effect).  We therefore 
divided the streamer tube length in 
six slices and computed the residuals in each slice 
separately. 
We generated a matrix of (14,2304,6) elements, where the first index
runs over the number of horizontal planes, the second over the wire number
and the last over the portion of the wire along its main axis.
We adopted an iterative
procedure, by adding at each step, for each element of the matrix, 
the mean value of the gaussian of the residuals belonging to each
portion of wire.  
As a results of this procedure,
Fig. 7 shows the distribution of the track residuals for the MACRO streamer
tube system in drift mode(black circles) and the MACRO simulation, GEANT
based, (continuous line).
The residuals of the down-going muons have a 
$\sigma=3 mm$, in good agreement with the MACRO 
simulation.  The continuous line shows
the residuals distribution for the streamer tube system in digital mode 
($\sigma=1 cm$), where we see
an improvement of the resolution by a factor $\simeq$ 3.5 has been obtained. 

For MACRO data however, we expect the resolution to be worse than that
measured in the PS-T9 test beam ($\sigma$=2mm) due to two effects.
From our simulation,
the most important contribution accounting for this difference, 
comes from $\delta$~-rays and 
radiated photons produced in the rock absorbers. 
Both of these effects spoil the space resolution by 
producing streamers closer to the wire than those coming 
from the muon, resulting in smaller drift radii. Moreover the MACRO
down-going muons, despite an average energy of 
$<$$E_{\mu}$$>$$\simeq$320 GeV still suffer
multiple scattering, mainly coming from the low energy tail of this 
distribution.

These hypotheses were tested during a second test beam,
performed at SPS-X7 in August 2001, where high energy muons with 
15~GeV $\leq$ E $\leq$ 100~GeV were available, with the same setup used
at PS-T9(Fig 8). 
The sigma of the residuals obtained with
$E_{\mu}$=100~GeV and rock absorbers inserted, was measured
$\sigma$=3~mm, in good agreement with that obtained using 
the MACRO down-going muon data.

\section{Muon energy estimate}
A muon energy estimate can be performed in MACRO by measuring the amount 
of muon multiple
scattering in the rock absorbers. The tests performed at CERN
PS/SPS beam lines, allowed us to demonstrate this as well as
offer the possibility of calibrating the MACRO system.
 
For each muon event we computed the following variables, sensitive to muon
multiple scattering.  The first three variables are just outputs from the
track fitting procedure:\\
1) The highest residual of the 14 measurements;\\
2) The average of the residuals; and\\
3) The standard deviation of the residuals.\\
For each track, we then considered the hit with the highest height and
that with the lowest height in the lower part of the detector
( i.e. excluding the Attico hits). Then we selected a median hit, having
the maximum distance in height from the other two hits. 
From this we constructed the next two variables:\\
4) The difference of the residuals of the highest hit and of the median
hit;~and\\
5)  The difference of the residuals of the lowest hit and of 
the median hit.

Lastly, we defined a ``progressive fit'' as the absolute value
of the residual $d_{i}$(i=1,14) as a function of the height of 
the streamer tube plane.
For a high energy muon, the average residual is roughly constant
in the different planes, since the muon energy is almost constant 
while crossing the experimental setup. For instance a 20 GeV muon
looses less than 5$\%$ of his energy after crossing the detector.
In contrast, a low energy muon
looses a high fraction of its energy, by ionization, crossing the rock 
absorbers. As a result,
the  average residuals are higher for the last crossed planes.
A linear fit of the absolute value of the residuals as a function
of the streamer tube number, gives a small
slope for high energy muons, while the slope is much larger
for low energy muons. Guided by this analysis we introduce the
following variables:\\
6) The slope of the ``progressive-fit''; and\\
7) The intercept of the ``progressive-fit''.

We followed a neural network approach(NN) in this analysis, 
choosing  JETNET 3.0\cite{jetnet},          
a standard package with a multilayer perceptron architecture and
with back-propagation updating. The NN was configured with 7
input variables quoted above and 1 hidden layer, selecting the 
Manhattan upgrading function.
Fig.~9 shows the distribution of the variables
quoted above and of the neural network output for muons with energy
$E_{\mu}$=100~GeV(continuous line) and for muons with energy 
$E_{\mu}$=2~GeV(dotted line).
Fig. 10 shows that the average neural network 
output increases as a function of the 
muon energy up to $E_{\mu}$$\simeq$40 GeV, saturating at higher 
energies.

The data collected during the PS test beam, provide an absolute energy 
calibration of the method, 
up to muon energy of 12 GeV. In order to check the neural network 
output in the whole energy range of Fig. 10, we used the data collected at the
CERN SPS-X7 beam line.

In Fig. 10 the test beam data and the Monte Carlo
prediction are compared: empty squares represent the Monte Carlo expectation,
black circles show the PS-T9 test beam points, while full triangles
are the SPS-X7 test beam data. 
The NN output obtained with the test beam data is properly  
reproduced by the Monte Carlo simulation.
The muon energy can be reconstructed by inverting the curve shown
in Fig. 10.  Fig. 11 and Table 1 show the reconstructed energy
for $E_{\mu}$=2,4,12,40 GeV: data collected at PS-T9 test beam(full
squares) and at SPS-X7 test beam (full triangles) are compared
with the Monte Carlo expectation(continuous line), showing a reasonable
agreement.
\begin{table}
\begin{tabular}{| c | c | c | c | c | c| }
\hline
$E_{\mu}$(GeV)&2. &3.&5.&10.&40. \\
\hline
&&&&&\\
Reconstructed energy  &($2^{+6}_{-1.5}$ )&($3^{+12}_{-2.5}$)&($5^{+18}_{-4}$) &($10^{+30}_{-8}$)&($40^{+60}_{-21}$)\\
(GeV)&&&&&\\
\hline
\end{tabular}
\caption{\label{tab:resol} Reconstructed muon energy}
\end{table}       

\section{Conclusions}
The use of the QTP-TDC's, offers the possibility of using the MACRO limited 
streamer tube
system in drift mode. The test beam run performed at CERN PS-T9 confirmed
such possibility.
The QTP system allows us to improve the 
streamer tube system space resolution by a factor of $\simeq$3.5,  
from $\sigma$$\simeq$$1 cm$ to $\sigma$$\simeq$3~mm. 
These improvements were realized by using a neural network 
approach in order to obtain an energy estimate of muons 
crossing the detector. The average neural network output increases as a function
of the muon energy up to $\simeq$40 GeV. The comparison between Monte
 Carlo expectation and the test beam data shows a good agreement. 
This method offers the possibility
to estimate the muon energy for neutrino induced 
upgoing muons in MACRO and thus to investigate the energy dependence of
the neutrino oscillation signal.

\vskip 1cm
{\bf Acknowledgements}

We would like to thank the CERN PS staff for the fruitful cooperation during
the test beam running.  We would like to especially thank 
R. Coccoli, Luc Durieu and T. Ruf for their help. We are 
also indebted to the efforts of
the CERN SPS staff and in particular to L. Gatignon and M. Hauschild
for their help on the preparation of the low energy beam we used. 

We gratefully acknowledge the support of the Director and of the staff of the Laboratori Nazionali del Gran Sasso and the invaluable assistance of the technical staff of the Institutions participating in the experiment. We thank the Istituto Nazionale di 
Fisica Nucleare (INFN), the U.S. Department of Energy and the U.S. National Science Foundation for their generous support of the MACRO experiment. We thank INFN, ICTP (Trieste), WorldLab and NATO for providing fellowships and grants (FAI) for non Italian 
citizens.

\newpage
\begin{figure}[ht]
 \begin{center}
  \mbox{\epsfig{file=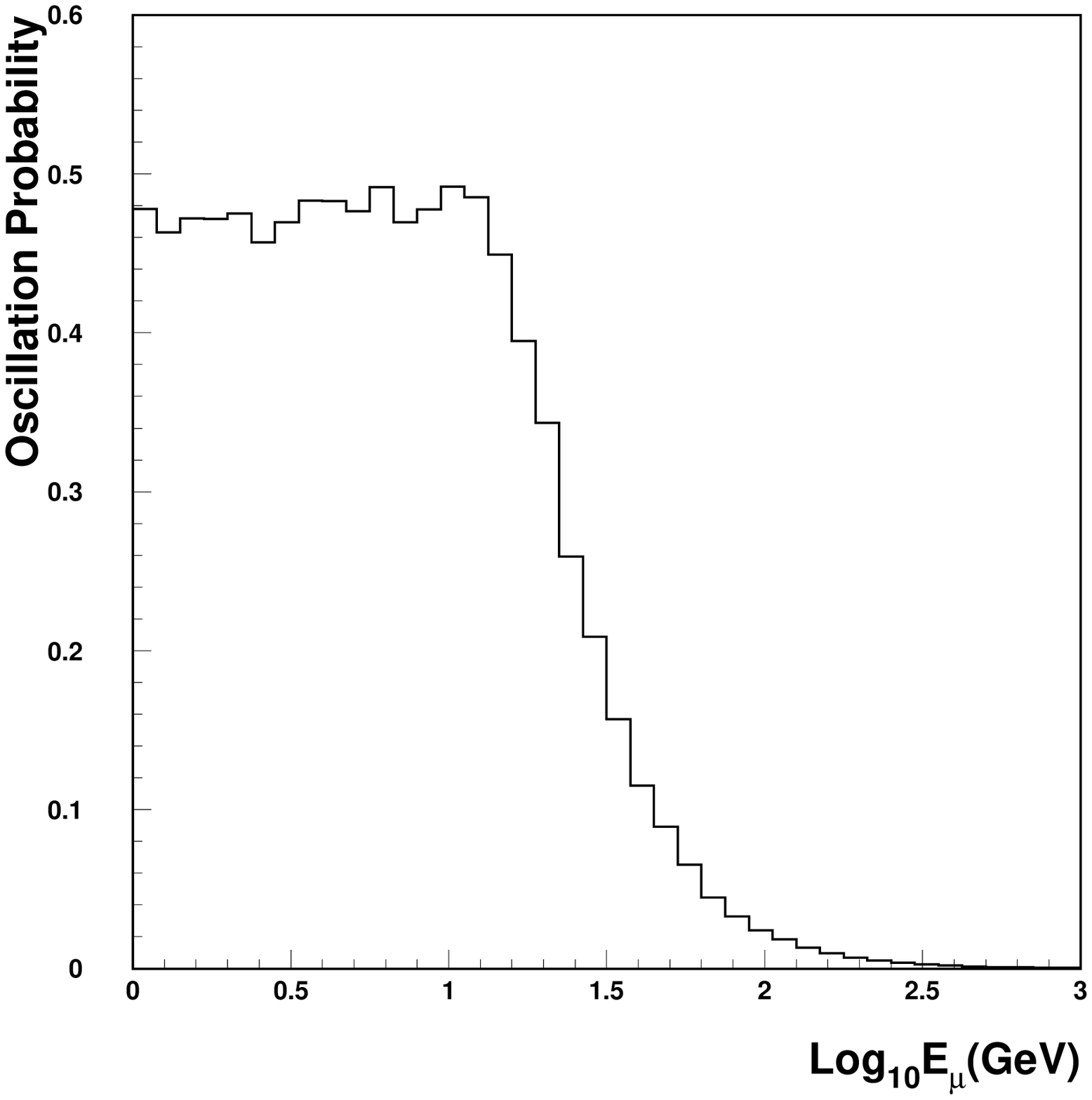,width=14cm}}
  \caption{\em Monte Carlo simulation: oscillation probability as a
  function of the energy
of the muon entering in MACRO for $\Delta m^{2}$=2.5$\cdot$$10^{-3}$$eV^{2}$,si$n^{2}$2$\theta$=1.
\label{fig:fig1}}
  \vspace{-0.5cm}
 \end{center}
\end{figure}   
\newpage


\begin{figure}[ht]
 \begin{center}
  \mbox{\epsfig{file=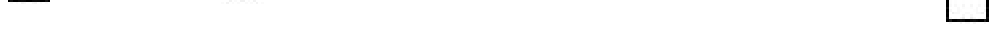,width=14cm}}
\vskip -14.5 cm
  \caption{\em Test beam layout at PS-T9: the trigger is provided by the fast
coincidence of the scintillators S1,S2,S3. \label{fig:fig2}}
  \vspace{-0.5cm}
 \end{center}
\end{figure} 
\newpage

\begin{figure}[ht]
 \begin{center}
  \mbox{\epsfig{file=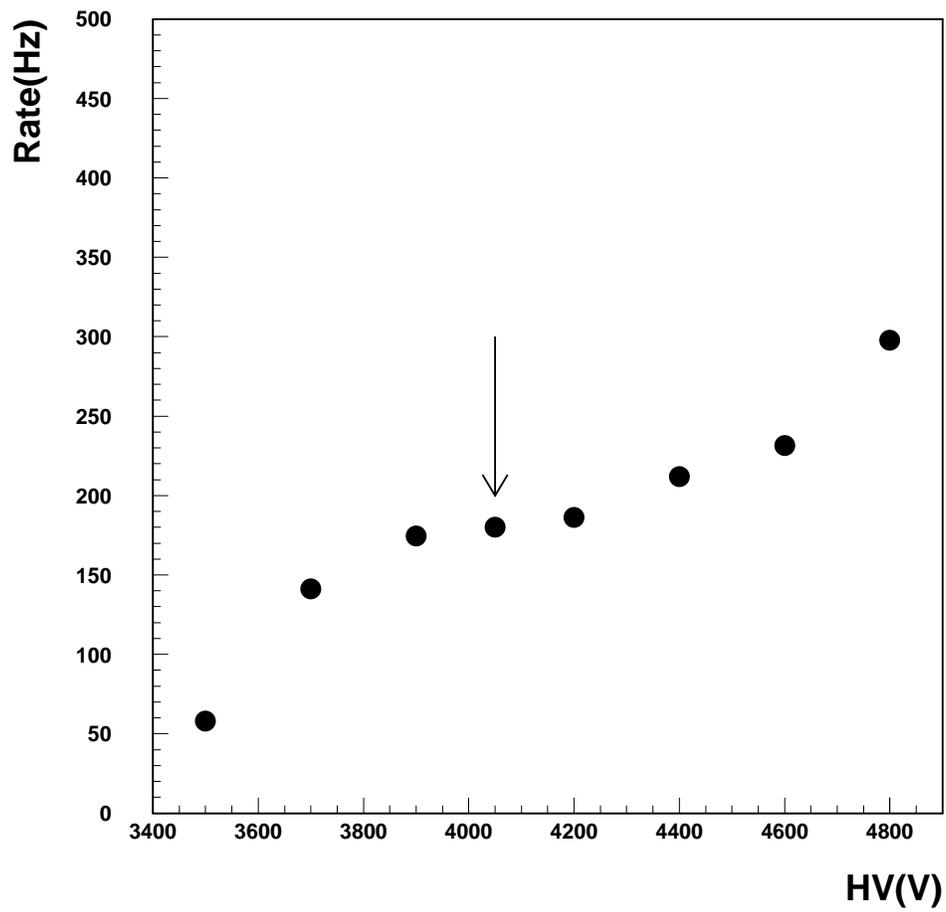,width=14cm}}
\vskip -3 cm
  \caption{\em Plateau of the streamer tubes: the arrow indicates the
working point. \label{fig:fig3}}
  \vspace{-0.5cm}
 \end{center}
\end{figure}

\newpage
\begin{figure}[ht]
 \begin{center}
  \mbox{\epsfig{file=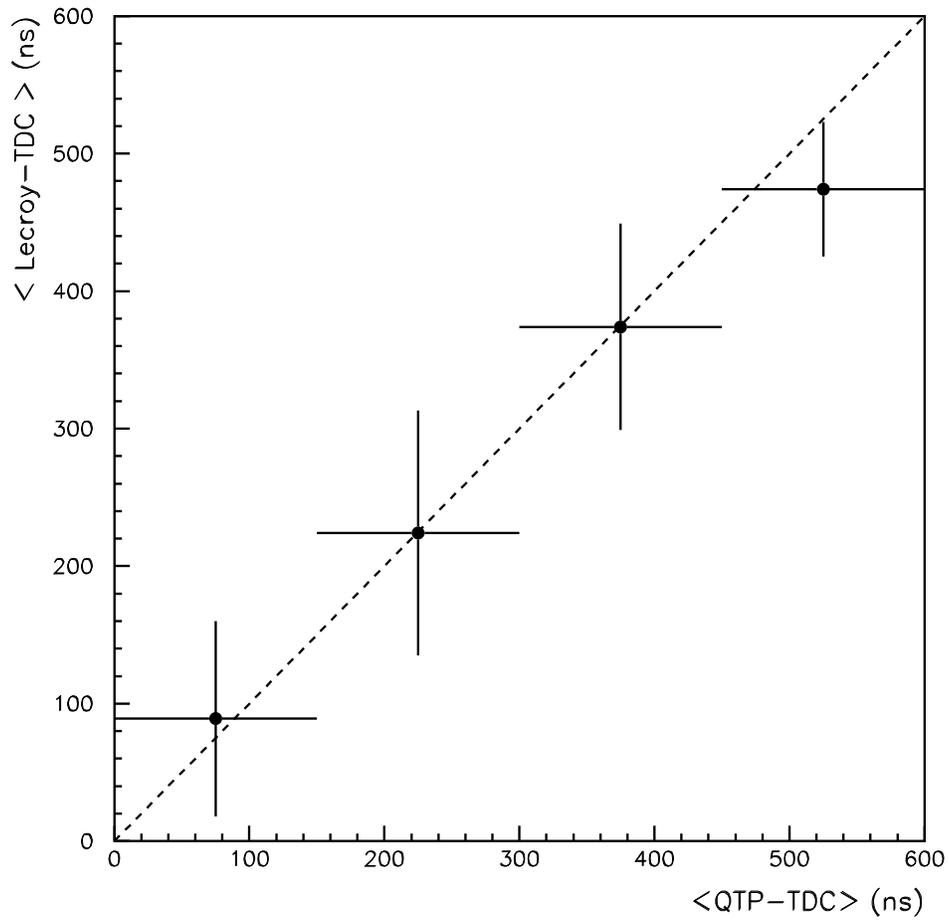,width=14cm}}
  \caption{\em  Profile plot of Lecroy 2228A TDC's as a 
function of QTP-TDC's. The $\simeq$10$\%$ shift of the last point, is due
 to the streamer tube electric field
  non-homogeneity
(see text). \label{fig:fig4}}
  \vspace{-0.5cm}
 \end{center}
\end{figure}      

\newpage
\begin{figure}[ht]
 \begin{center}
\vskip -3cm
  \mbox{\epsfig{file=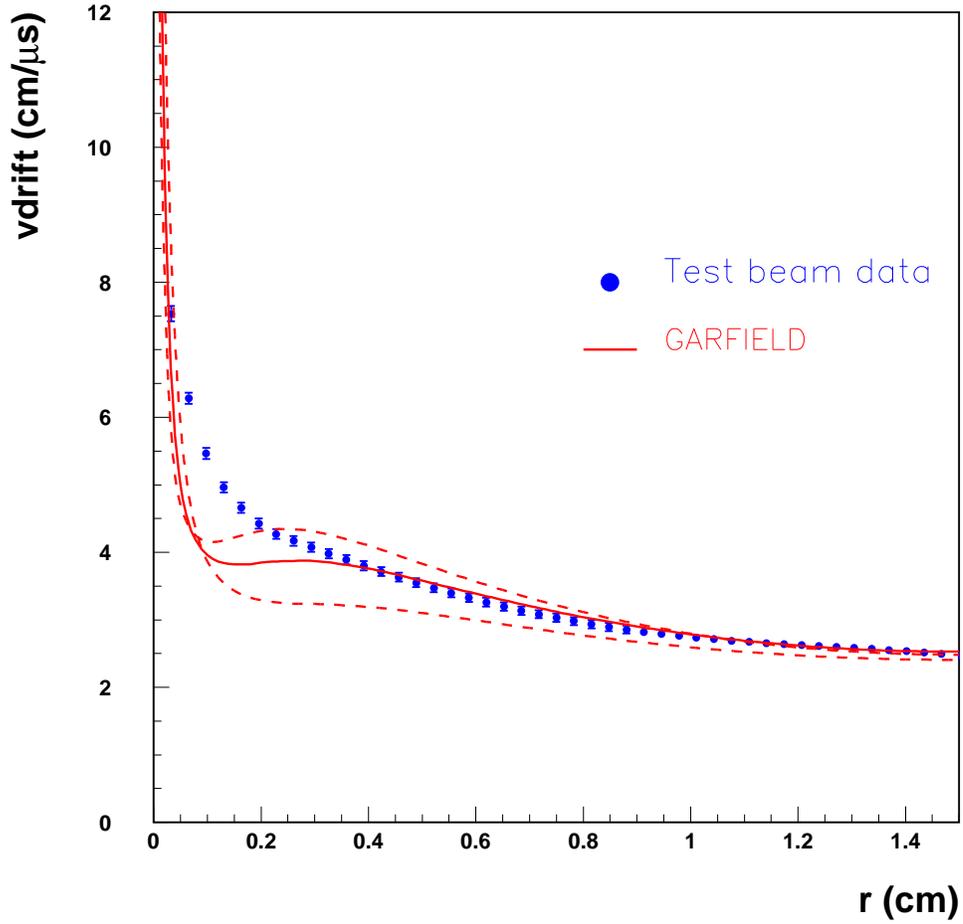,width=14cm}}
  \caption{\em  Drift velocity as a function of the distance from the wire,
measured at test beam and compared with the GARFIELD
  expectation. The dotted lines represent the effect of a 15$\%$ gas 
mixture variation.\label{fig:fig5}}
  \vspace{-0.5cm}
 \end{center}
\end{figure}      

\newpage
\begin{figure}[ht]
 \begin{center}
  \mbox{\epsfig{file=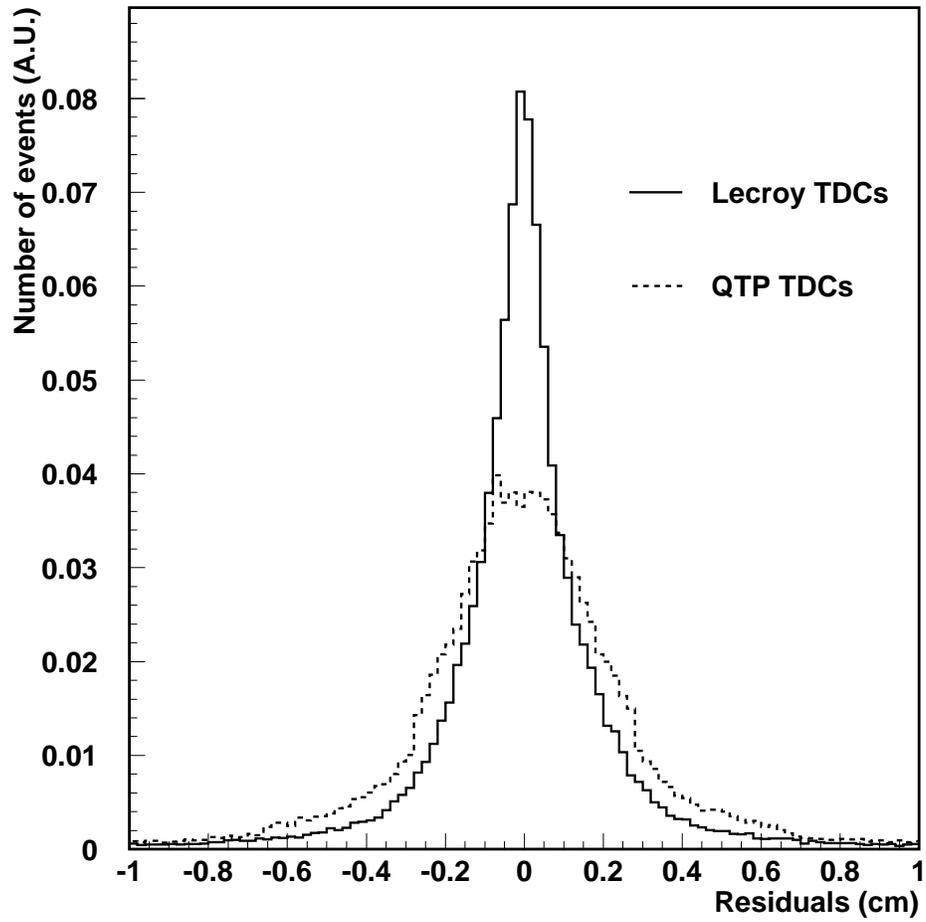,width=14cm}}
\vskip -4 cm
  \caption{\em  Test beam results: residuals distribution obtained using Lecroy 2228A TDC's
  (continuous line) and QTP-TDC's (dotted line).\label{fig:fig6}}
  \vspace{-0.5cm}
 \end{center}
\end{figure}      
\newpage

\begin{figure}[ht]
 \begin{center}
\vskip -3cm
  \mbox{\epsfig{file=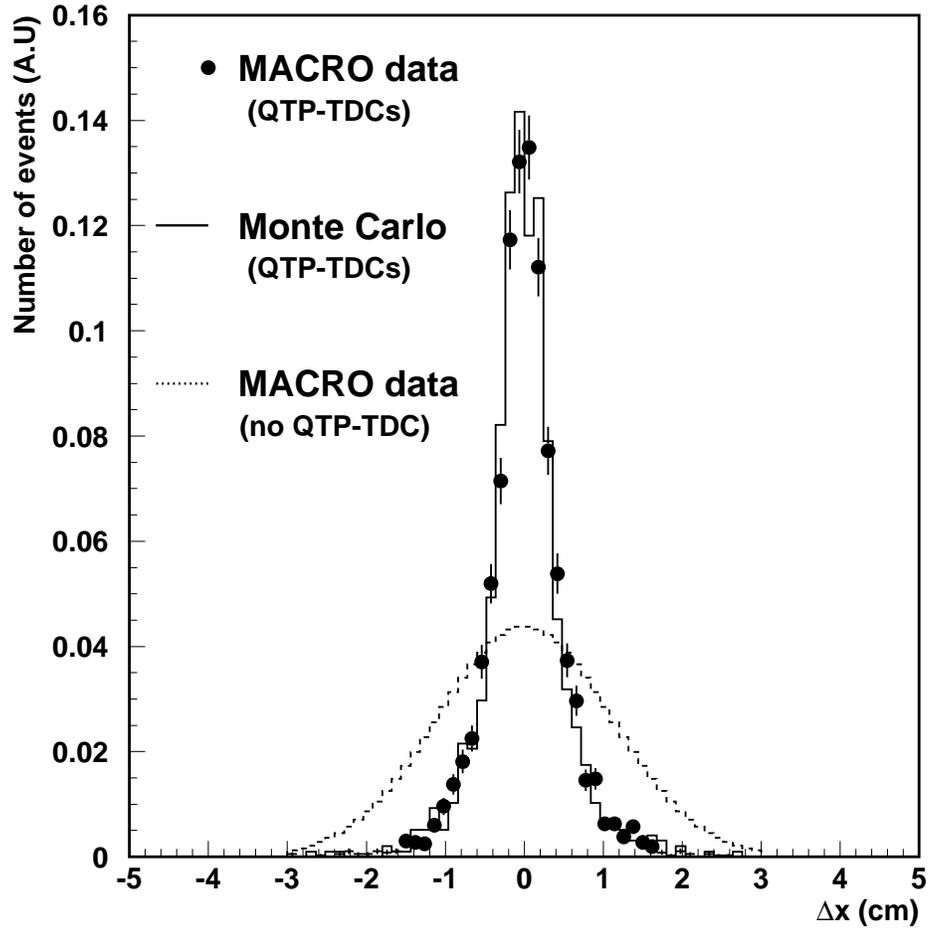,width=14cm}}
  \caption{\em  MACRO data: residuals distribution obtained using the MACRO QTP
TDC's($\sigma$=3 mm) compared with the Monte Carlo, GEANT based, expectation. The dotted 
line represents the residuals obtained using the MACRO streamer tube system 
in digital mode($\sigma$=1 cm).\label{fig:fig7}}
  \vspace{-0.5cm}
 \end{center}
\end{figure}
\newpage      
\begin{figure}[ht]
\begin{center}
\mbox{\epsfig{file=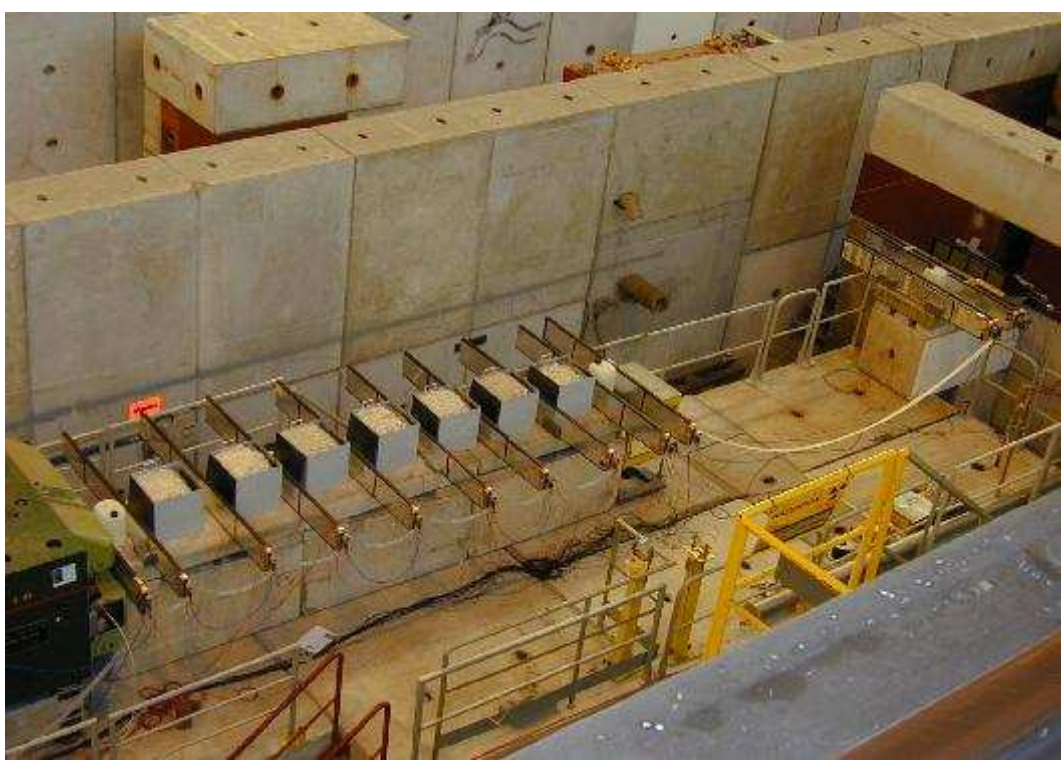,width=14cm}}
\caption{\em Photo of the test beam performed at CERN SPS-X7. \label{fig:fig8}}
\vspace{-0.5cm}
\end{center}
\end{figure}      

\newpage

\begin{figure}[ht]
\begin{center}
\vskip -5cm
\mbox{\epsfig{file=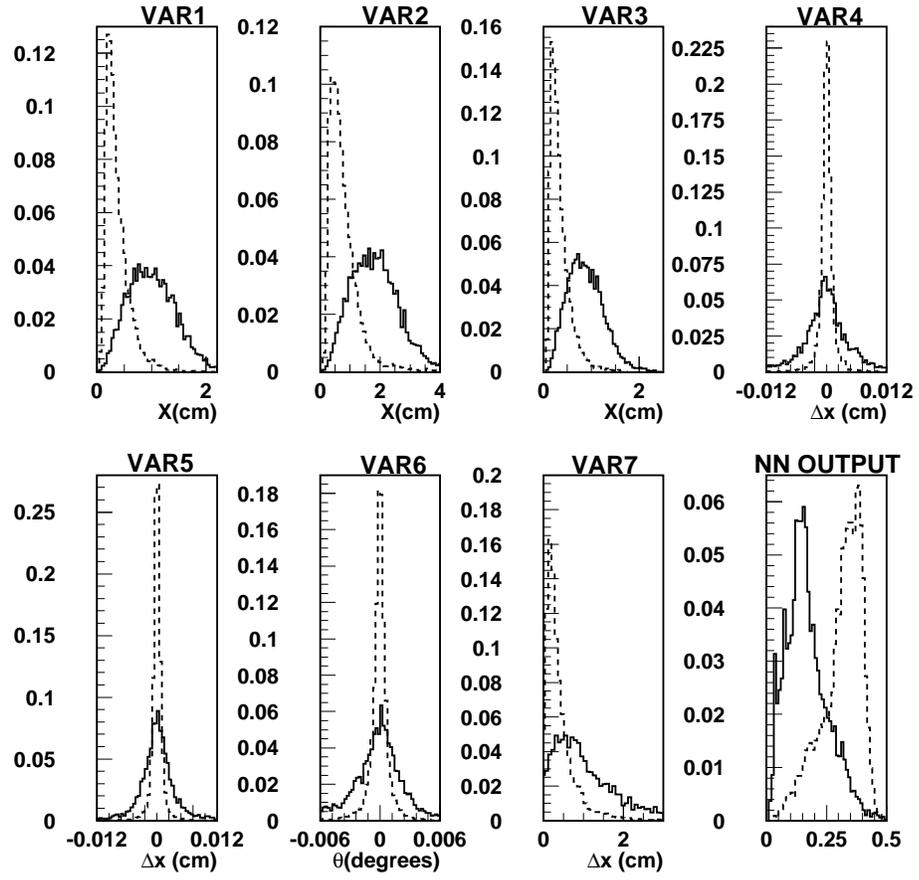,width=14cm}}
\vskip -4 cm
\caption{\em  Monte Carlo simulation: Distribution of the 7 input variables
and of the neural network output(continuous line $E_{\mu}$=2 GeV,dotted line
$E_{\mu}$=100 GeV)
.\label{fig:fig9}}
 \vspace{-0.5cm}
 \end{center}
\end{figure}      
\newpage
\begin{figure}[ht]
 \begin{center}
  \mbox{\epsfig{file=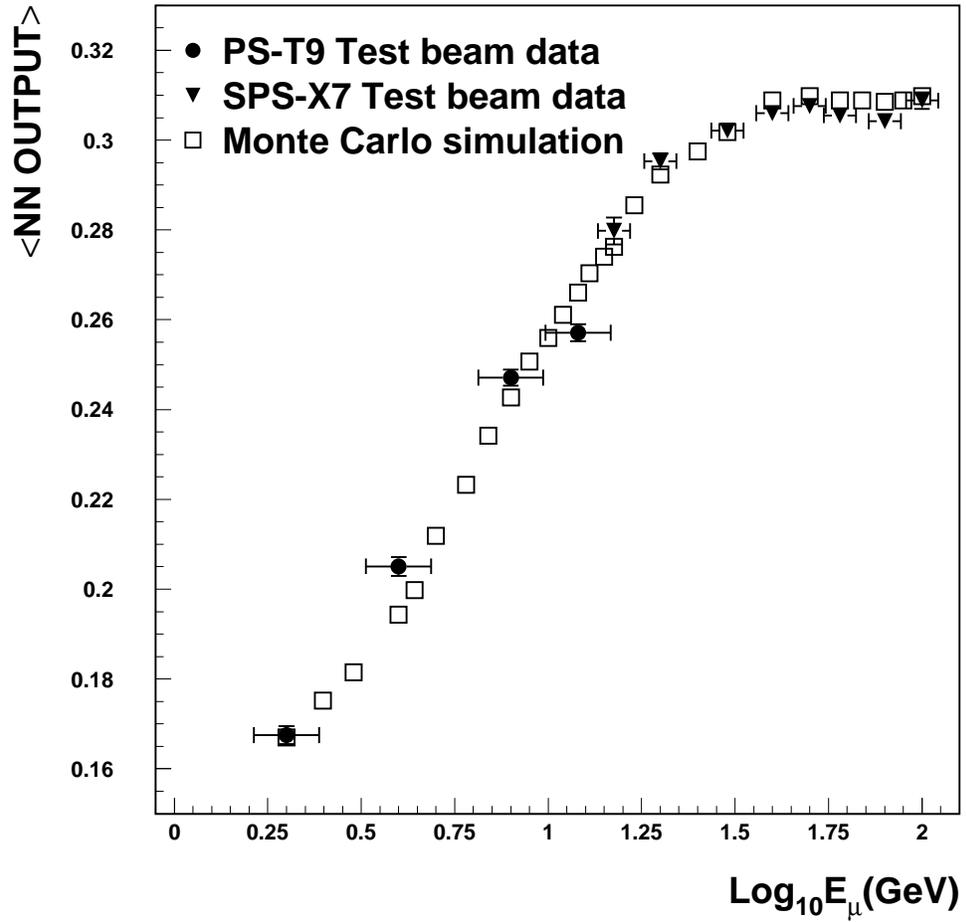,width=14cm}}
\vskip -4 cm
  \caption{\em Average neural network output as a
function of the muon energy: empty squares(Monte Carlo),
full circles (PS test beam data) and full
triangles (SPS test beam data).\label{fig:fig10}}
  \vspace{-0.5cm}
 \end{center}
\end{figure}      
\newpage
\begin{figure}[ht]
 \begin{center}
  \mbox{\epsfig{file=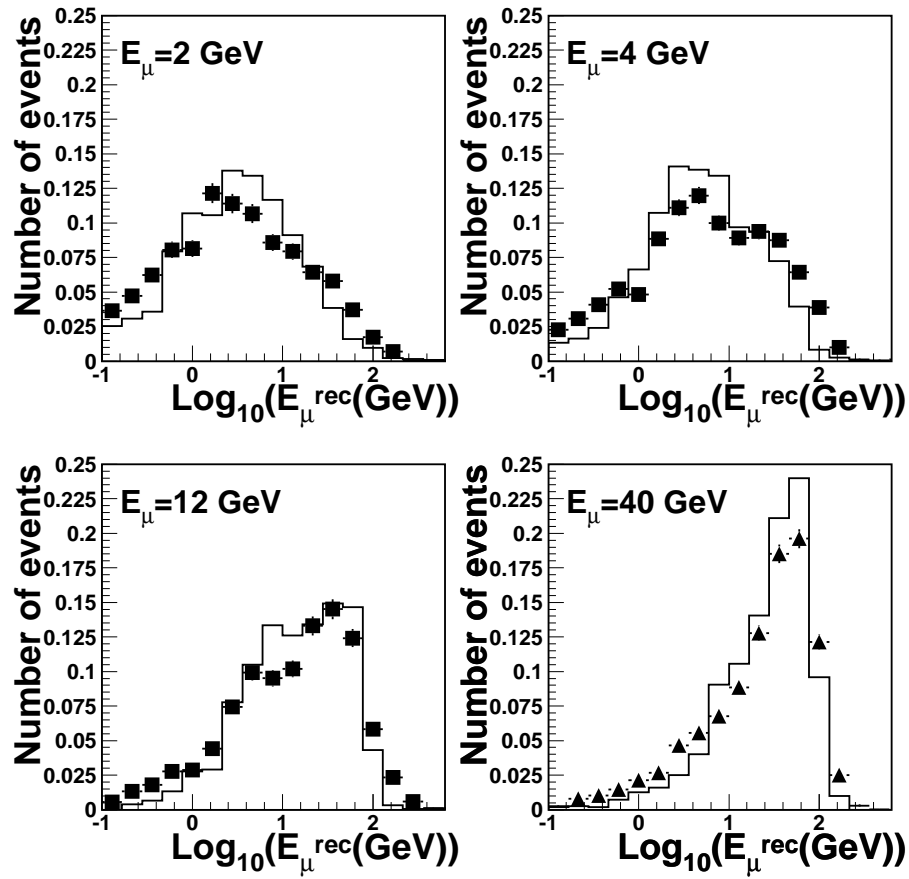,width=14cm}}
\vskip -4 cm
  \caption{\it  Reconstructed energy distribution for 2,4,12,40 GeV muons.
Monte Carlo: continuous line, test beam: full squares (PS-T9 data) and 
full triangles(SPS-X7 data).
\label{fig:fig11}}
  \vspace{-0.5cm}
 \end{center}
\end{figure}  

\end{document}